\documentstyle[aps,prl,twocolumn,epsf]{revtex}
     \begin{document}

     \draft
     \title{ Crossing of Specific Heat Curves in some correlated
     Fermion systems }
     \author{Suresh G. Mishra and P.A. Sreeram}
     \address{Institute of Physics, Sachivalaya Marg,
     Bhubaneswar 751 005, India.} 

     \maketitle

    \begin{abstract}
     Specific heat versus temperature curves for various pressures, 
     or magnetic fields (or some other external control parameter)
     have been seen to cross at a point or in a very small range of
     temperatures 
     in many correlated fermion systems. We show that this behavior 
     is related to the vicinity of a quantum critical point
     in these systems which leads to a crossover at some temperature $ 
     T^{\star} $ from
     quantum to classical fluctuation regime. The temperature at which 
     the curves cross turns out to be near $ T ^{\star} $.
     We have discussed the case of the normal
     phase of liquid Helium three and the heavy fermion systems 
     CeAl$_3$ and UBe$_{13}$ in detail within the spin fluctuation
     theory. When the crossover scale is any homogeneous function of these
     control parameters there is always crossing at a point.
    \end{abstract} 
    

     There has been a surge of interest in correlated fermionic
     systems for last ten years. This has led to a recognition that 
     the usual mean field or Hartree Fock description of interacting 
     fermionic systems is not enough, in particular when the effective 
     space dimension of the system is low or when the system is near 
     a quantum 
     phase transition due to the effects of characteristic low energy
     quantum fluctuations. For example, systems near a metal insulator
     transition or near a magnetic instability,  high temperature
     superconductors, heavy fermions and  liquid $^3$He, all
     show temperature dependence of their properties at low
     temperatures which differs from that expected in a normal Fermi
     liquid \cite{JPCM96}.     

     One phenomenon which had been observed long ago is that in some
     systems the specific heat curves as a function of temperature,
     for various values of external parameters (e.g. pressure,
     magnetic field) cross at a point or at least within a very
     narrow regime of temperature. This phenomenon was initially observed 
     for $^3$He by Brewer {\it et. al.} \cite{Brewer59} and has been
     seen later on,
     in a variety of materials ranging from systems close to metal-insulator
     transition to heavy fermions. The variety of
     materials in which this phenomenon has been observed leads one to 
     believe that there is some kind of universality in this
     behavior. In a recent publication, Vollhardt \cite{Voll97} has
     given a thermodynamic interpretation to this universality. The
     argument relies on a smooth crossover between behavior of entropy 
     at temperatures low compared to degeneracy temperature and the
     high temperature classical limit. As such, the question of why
     such crossings are prominently seen in systems with highly
     enhanced magnetic susceptibility or effective mass remains
     unanswered. Here we propose that the operative cause is the
     proximity to a quantum critical point (or $ T = 0 $ critical
     point). Vicinity to a quantum 
     critical point is usually marked by enhancement in the effective mass, 
     and in spin or density (charge) response in a system at low
     temperatures. This in
     turn introduces a low energy scale which marks a crossover from
     quantum to a classical behavior in the temperature dependence of
     various physical properties. In most materials the abovementioned 
     crossing of specific heat occurs near this crossover
     temperature. This scenario is quite general and
     holds for transitions involving conserved (for example, the
     ferromagnetic) as well as nonconserved (the 
     anti-ferromagnetic) order parameters. The examples discussed in
     the present letter   
     have been chosen to represent both of these order parameter
     fluctuations. We use the microscopic spin fluctuation theory
     \cite{MR78,MOR85} to discuss the behavior in detail. This theory
     has the
     low energy scale inherently built in it.

     Consider first the case of liquid $^3$He. It is a Fermi system 
     with a degeneracy temperature of about 5 K. It has some
     interesting normal state properties. For example, it behaves like
     a dense classical liquid down to 0.5 K and like a degenerate Fermi 
     liquid below 0.2 K. It has a large (nuclear) spin susceptibility, 
     about 10 to 25 times the free Fermi gas or Pauli susceptibility
     $\chi_P$, 
     depending on pressure. The coefficient of the
     linear term in specific heat is also large. In the spin
     fluctuation theory presented below, the liquid is regarded as near a 
     ferromagnetic instability. In this theory the temperature
     variation of various physical
     quantities is governed by transverse and longitudinal spin
     fluctuations. Though the actual transition does not take place,
     the effect of fluctuations is observable over a wide temperature
     range at low temperatures.\cite{MR78,MR85}. 

     In the following we use some results from our earlier works
     \cite{MR78,MR85,SGM98} to
     discuss the crossing point in the specific heat  
     curves. We consider first fluctuation above the transition to a
     ferromagnetic phase.
     The spin fluctuation contribution to the free energy within the
     mean fluctuation field approximation (or quasi harmonic
     approximation) at temperature $T$ for systems near a
     ferromagnetic instability is given by \cite{MR85},
     \begin{equation}
     \Delta\Omega =\frac{3 T}{2}\sum_{q,m}\ln \{ 1-U\chi_{qm}^0
     +\lambda T \sum_{q^\prime,m^\prime}D_{q^\prime m^\prime} \}.
     \label{eqn:freeenegy1}
     \end{equation}
     Here $D_{q,m}$ is the fluctuation propagator which is related
     to inverse dynamical susceptibility, $\chi_{qm}^0$ is the
     free Fermi gas (Lindhardt) response function, and $\lambda$ is the
     fluctuation coupling constant. Considering only the
     thermal part of the integral and ignoring the zero point part,
     we perform the frequency summation and obtain, 
     \begin{equation}
     \Delta\Omega_{Th}=\frac{3}{\pi}\sum_q \int_0^\infty 
     \frac{d\omega}{e^{\omega/\tau}-1} \arctan \{\frac{\pi\omega/4q}
     {\alpha(\tau )+\delta q^2}\},
     \label{eqn:freeenegy2}
     \end{equation}
     where  $\alpha (\tau ) $ is the inverse of spin
     susceptibility in units of the Pauli susceptibility.  The
     wavevector $ q $ is given in units of 
     Fermi momentum $ k_F $ and the energy is in units of Fermi
     energy ($ \tau = T/T_F$). For a free Fermi gas $ \gamma = 1/2 $, 
     $\delta = 1/12 $. The correction to the specific heat is given by,
  \begin{equation} 
     \frac{\Delta C}{k_B}  = - 3\tau^2\sum_q
     \biggl[(\frac{2}{\tau}\frac{\partial y}{\partial
     \tau}
     +\frac{\partial^2 y}{\partial \tau^2})\phi(y)+(\frac{\partial
     y}{\partial \tau})^2
     \frac{\partial\phi(y)}{\partial y} \biggr] \label{eqn:ferro3dspe1}
     \end{equation} 
    The function $ \phi (y) $ is given by $ \ln y - 1/2y - \psi (y)
     $ and $ \psi (y) $ is digamma function  where, $ y = q
     (\alpha(\tau) + \delta q^2) / (\pi^2 \gamma \tau) $. $ \phi (y) $ 
     is  related to the
     fluctuation self energy summed over frequency. It varies as $
     1/2y $ for $ y \ll 1 $ and as $ 1/12y^2 $ for $ y \gg 1 $.
    
      Clearly the calculation of specific heat correction involves the
     temperature dependence of spin susceptibility. A self consistent
     equation for the temperature dependence of $ \alpha (T) $ 
     within one spin
     fluctuation approximation is given by \cite{MR78,SGM98},
     \begin{equation}
     \alpha(\tau) = \alpha(0) + \frac{\lambda}{\pi}\sum_q 
     q \phi(y).
     \label{eqn:ferroalpha1}
     \end{equation}
     For a finite $\alpha(0)$ there are two regions of temperature
     \cite{MR78}. For $ \tau < \alpha(0)$, which corresponds to $ y
     \gg 1 $, one gets an enhanced Pauli susceptibility with standard
     paramagnon theory corrections, 
     $ \alpha(\tau ) = \alpha (0) + a \tau^2 / \alpha (0) $, 
     where $a$ turns out to be $0.44$. At higher temperatures (  
     $ \alpha(0) < \tau < 1 $) $ \alpha (\tau ) \sim \tau^{n} $ with
     the exponent $ 1 \le n \le 4/3 $. This result for the
     susceptibility mimics the classical Curie Weiss behavior.
     Notice that even in a degenerate regime ($ \tau < 1 $),
     the susceptibility for a Fermi system behaves like the one for 
     a collection of
     classical spins. This behavior agrees well \cite{MR78} with
     experimental results of Thompson et. al. \cite{Thompson70}.
     The parameter $\alpha (0) T_F $ is the low energy scale which
     arises in the spin fluctuation theory naturally. 
     The corresponding low temperature ($\tau \le \alpha (0)$)
     correction to the specific heat is,
     \begin{equation}
     \frac {\Delta C }{k_B} = - \sum_q \frac {\pi^2 \tau } {4
     q(\alpha + \delta q^2 )}.
     \label{eq:lowtempspe1} 
     \end{equation}
     The phase space integral reproduces the standard paramagnon mass
     enhancement result, $ \tau \ln \alpha $ for $ \Delta C $.           
     In the classical regime, $\alpha (0) \le \tau \ll 1 $, where the small
     $y$ approximation holds and $ \alpha (\tau ) $ varies as $ \tau,
     $  $ \Delta C $ falls as $ 1/\tau^2 $ and vanishes at higher 
     temperatures.  

     The main point of the above discussion is that there are two
     regimes for specific heat
     similar to the regimes in the susceptibility variation. The
     behavior of the specific heat in these two regimes is
     qualitatively different. At low temperature there is an enhanced
     linear rise of specific heat correction with temperature leading
     to a peak  
     and thereafter a slow fall as the temperature increases. The
     peak marks a transition from quantum to classical spin
     fluctuation regimes.  
     Considered as a function of $\alpha (0) T_F$, the temperature
     dependence of 
     specific heat is more revealing. Below a certain temperature $T_{cr}$,
     specific heat decreases as $\alpha (0) T_F$ increases, while above
     it the behavior is reversed (see Fig. \ref{fig:cvvsal}). $ T_{cr}
     $ clearly marks the
     crossing and is of the order of $ \alpha (0) T_F $. The spin
     fluctuation theory has only one parameter, 
     that is, $\alpha (0) T_F$. The pressure or magnetic field dependence
     of quantities is realised through the dependence of $\alpha (0)
     T_F$ on them. Whenever $ \alpha (0) T_F$ is homogeneously increasing or
     decreasing function of these parameters the specific heat curves
     will cross at a point. In this case $ \partial C / \partial
     \alpha (0)T_F = 0 $ 
     at  $ T = T_{cr} $ also means $ \partial C / \partial X =
     0 $ at the same temperature, where $X$ is an external control
     parameter like pressure or magnetic field. The later equation is
     the condition for crossing of curves at a point.

     For liquid $^3$He the specific heat is plotted in
     Fig. \ref{fig:thhe3} as a function of temperature for various
     values of pressure, assuming a linear reduction of $\alpha (0)
     T_F $ with pressure. The linear scaling is  
     experimentally observed above pressures about 15 kbar. However,
     at small pressures there is some departure. The peak in
     $\bigtriangleup C(T)$ appears around $0.15K$. To
     calculate the specific heat, the free Fermi gas part 
     ($\pi^2 T/2 T_F$) has been added to $\Delta C(T)$. The value
     of $\alpha(\tau)$ has been calculated self consistently using
     Eq.(\ref{eqn:ferroalpha1}) and then used as an input in the specific
     heat calculation. The coupling constant $\lambda$ has been
     chosen to be $0.08$ and the cutoff for the momentum sum, 
     $1.2k_F$. The crossing temperature is related to 
     $ \alpha (0) T_F $ which depends on pressure in general.
     The crossing point shifts towards high temperature side slightly 
     with increase in cutoff and with decrease in $\lambda$ but the nature
     of crossing is not affected. 
     
     There are some heavy fermion materials in which the specific
     heat curves cross. We consider the case of CeAl$_3$
     \cite{Brodale86} and UBe$_{13}$ \cite{Phillips87}. CeAl$_3$ does
     not undergo either a magnetic or a superconducting transition,
     while UBe$_{13} $ becomes superconductor at 0.9 K at normal
     pressure. The present discussion pertains 
     to their normal state properties only. Heavy fermions are
     characterised by a large linear temperature dependent term in the
     specific heat and a large low temperature spin susceptibility
     \cite{Grewe91}. In this regime the resistivity also shows a T$^2$ 
     behavior characteristic of a Fermi liquid. Above a certain
     temperature $T^{\star}$, the susceptibility starts showing a
     Curie Weiss behavior, indicating the existence of interacting
     local moments on the f-shells. The local moment to Pauli like
     behavior of the susceptibility, as temperature reduces, marks the
     onset of coherence in these systems. In UBe$_{13}$ this coherence
     regime is less visible because of the onset of superconductivity,
     but once the superconductivity is suppressed on application of
     pressure the coherence is restored \cite{Aronson89}. At present a
     clear microscopic understanding of  
     the behavior of heavy fermions is lacking, one has to take
     recourse to various levels of phenomenology. It is possible that
     the unusual low temperature dependence of physical properties in
     UBe$_{13}$ for example, is due to its being a non-fermi liquid
     of as yet unknown origin. We take the point of view here that
     this behavior can be described in terms of proximity to a quantum 
     critical point which is also known to lead to temperature
     dependences different from fermi liquid theory (for example
     \cite{SGM98}). 
 
     Because of the similarity to liquid $^3$He, at the
     phenomenological level it is tempting to apply the spin  
     fluctuation theory to these materials, with $\alpha (0) T_F $ 
     playing the role of the crossover temperature $T^{\star}$.   
     However, there is a difference. While $^3$He can be considered
     close to a ferromagnetic transition, most heavy fermion materials seem
     close to an antiferromagnetic instability. In the present work,
     we therefore consider the heavy fermions in the coherence regime
     as nearly antiferromagnetic Fermi liquid.
     
     We have calculated the specific heat corrections by writing the
     equations for the susceptibility enhancement and specific heat
     near an antiferromagnetic instability. The formalism remains same
     except that the factor $\omega/q$ in Eq. \ref{eqn:freeenegy2} is
     replaced by $\omega$ to take care of low energy behavior of the
     fluctuation propagator \cite{SGM98}. The difference is due to the
     fact that the order parameter does not remain a conserved quantity. 
     Further, to reproduce the huge effective mass observed,
     fluctuation modes are essentially dispersionless in heavy
     fermions \cite{Conti98}, namely the coefficient of the $ q^2 $
     term in $ y $, i.e., $\delta \approx 0$. In this case, the
     leading contribution to specific heat is $\tau/\alpha(0)$ at low
     temperatures. In the same range of temperatures the leading
     temperature correction to zero temperature susceptibility is 
     $\tau^2/\alpha^2(0)$. 

      In Fig-\ref{fig:ceal3} and \ref{fig:ube13} the specific heat
      curves for CeAl$_3$ and UBe$_{13}$ have been plotted as a
      function of temperature for 
      various pressures. The value of $\gamma$ has been taken to 
      be 0.185 and the cutoff $q_c$ is 2.0. The fluctuation coupling
      $\lambda$ is 5 x 10$^{-4}$ for CeAl$_3$ and 2 x 10$^{-4}$ for
      UBe$_{13}$, and decreases slightly with pressure. 
      The parameter $\alpha (0) T_F$ is of the order of 
      the crossing temperature with a weak linear pressure dependence.
      The variation with pressure is within 10 $\%$. In contrast to
      $^3$He, here $\alpha (0) T_F $ increases with pressure. This is
      because in $^3$He pressure brings the atoms closer and
      thereby increasing the interaction, while in heavy fermions the
      reduction in the lattice parameter enhances the hybridization
      between conduction electrons and f electrons thereby the
      antiferromagnetic exchange between local moment and the
      conduction electron will be enhanced leading to a non magnetic
      ground state. It is seen that the curves cross 
      within a small regime close to the experimental crossing point.
      Beyond the crossing point the deviation from xthe experimental
      curves is large. In fact, in heavy fermions, the
      curves cross at two points, the second point being away from 
      the crossover temperature T$^\star$, though still at
      temperatures far below $ T_F$. The reason for the second
      crossing cannot be found in a single parameter theory like the
      present one. It might be due to some other low lying modes like
      crystal field excitations or phonons \cite{Voll97}.

      So far we have discussed the ferro- and antiferromagnetic quantum 
      critical points. In a phenomenological model attempting to
      incorporate some aspects of strong correlations near the Mott
      transition Rice et.al. generalized the Brinkman-Rice theory to
      finite temperatures by introducing an extra ansatz for the
      entropy. It was applied to the case of UBe$_{13}$
      \cite{Rice85} and later to liquid $^3$He \cite{Seil86}. At a
      low energy scale which is related to reducing double occupancy
      there is crossover between Pauli to Curie behavior for the
      susceptibility. However, the specific heat curves \cite{Seil86}
      for liquid $^3$He at various pressures seem to cross over a wide
      range of temperatures unlike the experimental findings
      \cite{Grey83}. Recently, the metal insulator transition has been
      discussed within the single band Hubbard model for infinite
      dimension by Georges and Krauth \cite{Georges93}. A low energy
      scale, related to the vanishing quasiparticle weight, 
      arises in the metallic side of the transition. The specific heat
      curves cross at temperature around this scale. However, the
      theory gives a second crossing around the energy scale
      $ U $.  

     We have used the terms quantum and classical in the discussion
     above, because, the temperatures below $\alpha (0) T_F $ essentially
     define a regime where one gets a Fermi liquid behavior whereas
     at high temperatures, fluctuations get correlated resulting in
     the classical behavior for the susceptibility.  The distinction,
     quantum versus classical, becomes clear when one takes the limit
     $ \alpha (0) \rightarrow 0 $ (the quantum critical point). In
     that case the Curie law for susceptibility is
     obtained down to zero degree \cite{SGM98}, while in the opposite
     limit ($\alpha (0) \rightarrow 1 $) one gets the Pauli
     susceptibility; in either of these limits the curves for specific 
     heat do not cross. 
    
     Acknowledgement : We are grateful to Prof. T. V. Ramakrishnan for 
     critical reading of the manuscript.

     \begin{figure}
     \caption{Specific Heat as a function of $\alpha(0)T_F$ for 
      CeAl$_3$ (to be discussed later in the text) for various 
      temperatures calculated from the spin fluctuation theory. A
      similar behavior is obtained for $^3$He.}        
     \label{fig:cvvsal}
     \end{figure}

     \begin{figure}
     \caption {$C(P,T)$ for $^3$He with 
      $\alpha (0) T_F$ assumed to vary linearly with pressure.}
     \label{fig:thhe3}
     \end{figure}
      
      \begin{figure}
      \caption {Semilog plot of $C(P,T)/T$ as a function of $T$ for
      $CeAl_3$ for variuos pressures. The symbols are experimental
      points (Ref. 12) and the lines are results from the 
      spin fluctuation theory.}
      \label{fig:ceal3}
      \end{figure}
      
      \begin{figure}
      \caption {Semilog plot of $C(P,T)/T$ as a function of $T$ for
      UBe$_{13}$, above the superconducting transition temperature,
      for various pressures. The symbols are experimental
      points (Ref. 12) and the lines are results from the 
      spin fluctuation theory.}
      \label{fig:ube13}
      \end{figure}


\vfill

      \begin{thebibliography}{99}
      \bibitem{JPCM96} See for example, J. Phys.: Condens. Matter {\bf
          8}(48) (1996), special issue on non-Fermi-liquid behavior in 
        metals, edited by P. Coleman, B. Maple, and A. Millis.
      \bibitem{Brewer59} D. F. Brewer, J. G. Daunt and
        A. K. Sreerdhar, Phys. Rev. {\bf 115} (1959) 836.
      \bibitem{Voll97} D. Vollhardt, Phys. Rev. Letts.
        {\bf 78} 1307 (1997).
      \bibitem{MR78} S. G. Mishra and T. V. Ramakrishnan, Phys Rev.
        {\bf B18}, 2308 (1978).  
      \bibitem{MOR85} T. Moriya, {\it Spin Fluctuation in Itinerant 
       Electron Magnetism} (Springer, Heidelberg, 1985).
      \bibitem{MR85} S. G. Mishra and T.V.
        Ramakrishnan, Phys. Rev.  {\bf B31} 2825 (1985).  
      \bibitem{SGM98} S. G. Mishra and P. A. Sreeram, Phys. Rev. B
        {\bf 57} 2188 (1998).       
      \bibitem{Thompson70} J. R. Thompson, Jr., H. Ramm, J. F. Jarvis
        and  Horst Meyer, Jour. Low. Temp. Phys., {\bf 2} 521,539 (1970).  
      
      \bibitem{Brodale86} G.E. Brodale, R.A. Fisher, N.E. Phillips and 
        J.  Flouquet, Phys. Rev. Lett. {\bf 56} 390 (1986). 
      \bibitem{Phillips87} N.E. Phillips, R.A. Fisher, J. Flouquet,
      A.L. Giorgi, J.A. Olsen and G.R. Stewart, J. Mag. Magnetic Materials 
      {\bf 63 \& 64} 332 (1987). 
      \bibitem{Grewe91} For a review see N. Grewe and F. Steglich,
      {\it Handbook on the Physics and Chemistry of Rare Earths} Vol 
      14, Eds. K.A. Gschneidner Jr., and L. Eyring {Amstardam
      Elsevier} p. 343 (1991).
      \bibitem{Aronson89} M. C. Aronson, J. D. Thompson, J. L. Smith,
        Z. Fisk and M. W. McElfresh, Phys. Rev. Lett. {\bf 63} 2311 (1989).
      \bibitem{Conti98} M.A.  Continentino, Phys. Rev. {\bf B 57} 5966
       (1998).
      \bibitem{Rice85} T.M. Rice, K. Ueda, H.R. Ott, and H. Rudigier,
      Phys. Rev.{\bf B 31} 594 (1985).
      \bibitem{Seil86} K. Seiler, C. Gros, T.M. Rice, K. Ueda, and 
       D. Vollhardt, J. Low Temp. Phys. {\bf 64} 195 (1986).        
     \bibitem{Grey83} D.S. Greywall, Phys. Rev. {\bf B27} 2747
       (1983). 
      \bibitem{Georges93} A. Georges and W. Krauth, Phys.Rev. {\bf B
          48} 7167.    

     \end{thebibliography}
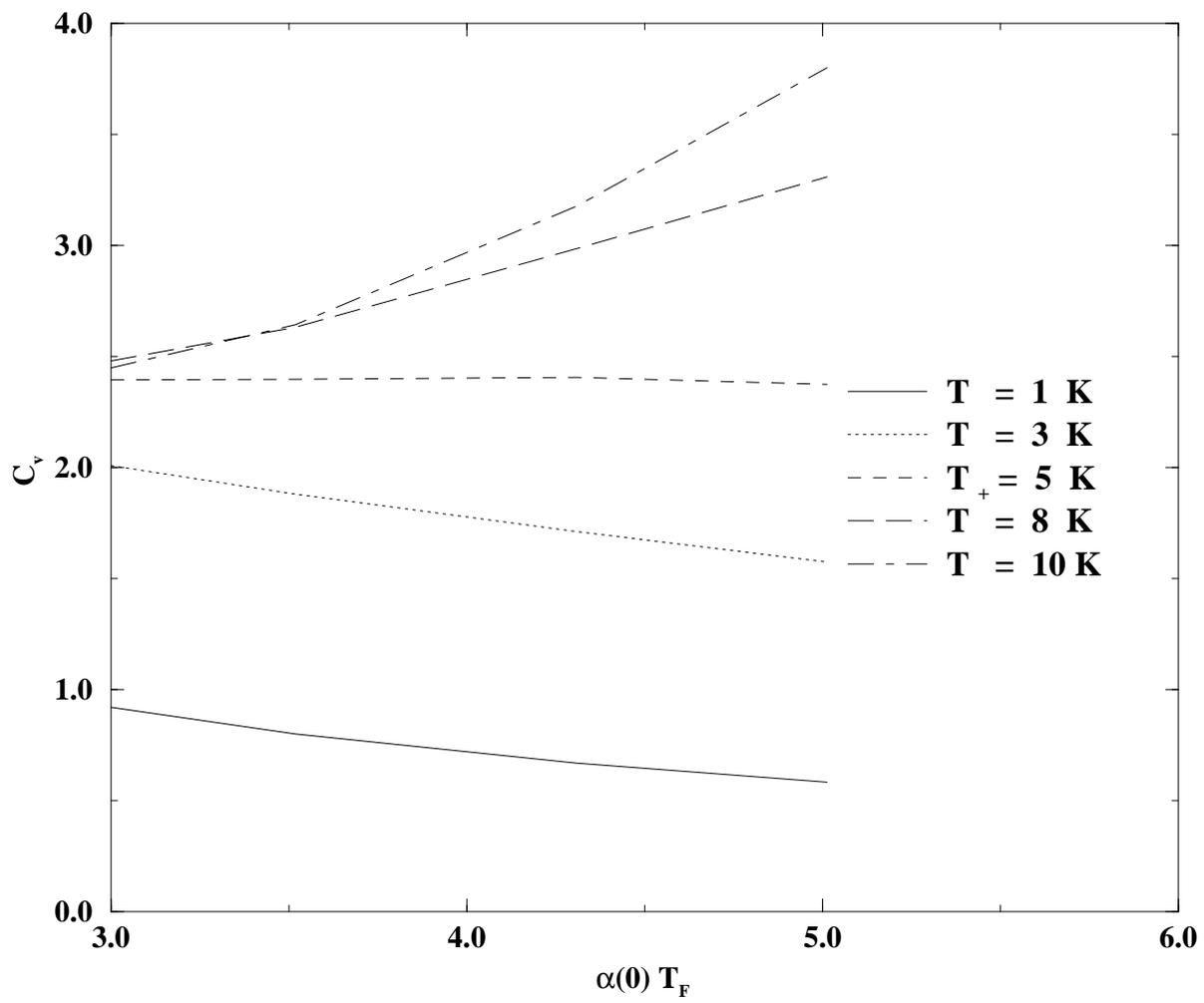
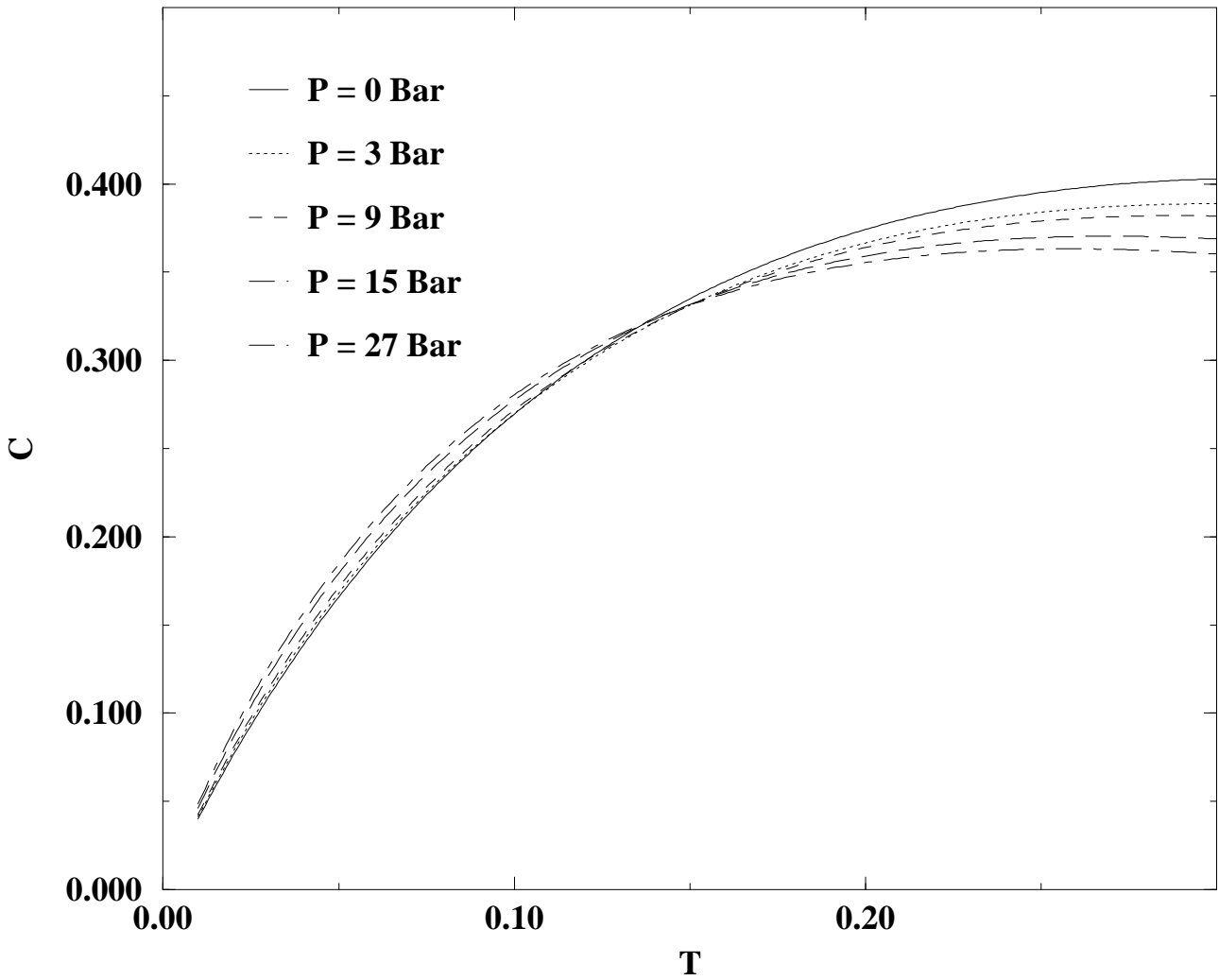
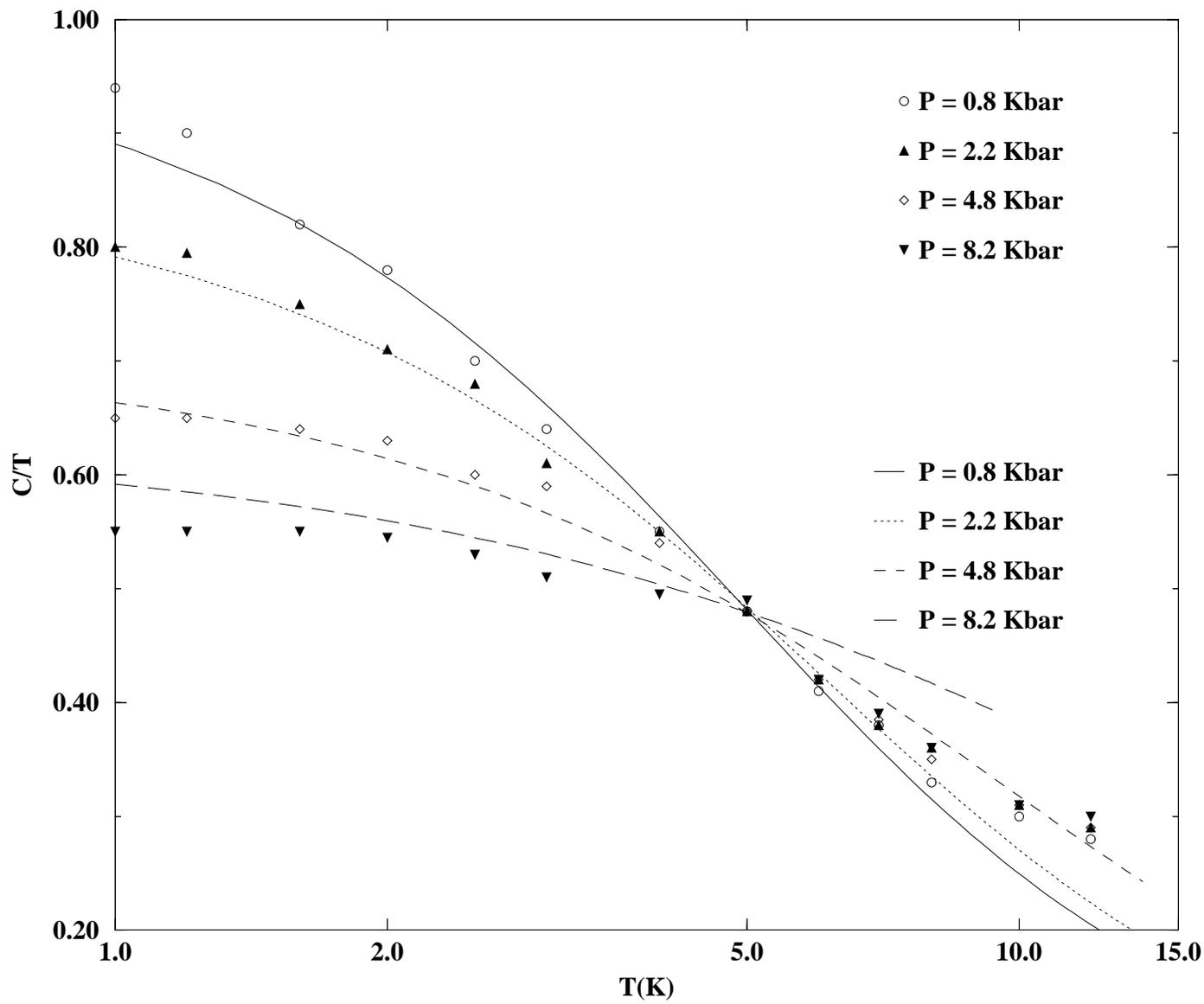
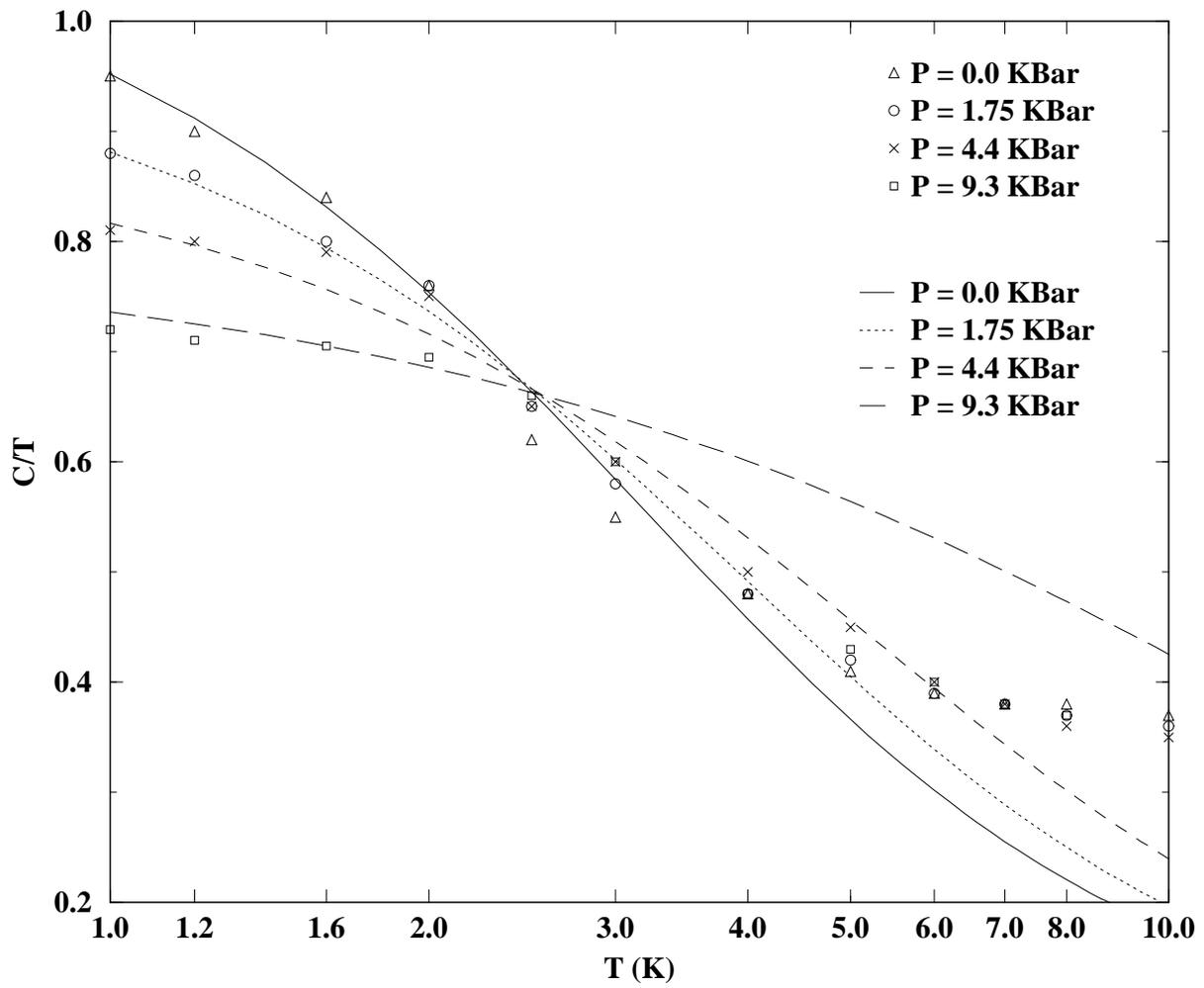
\end{document}